\documentclass[12pt]{article}
\usepackage{amsmath}
\def\be{\begin{equation}}
\def\ee{\end{equation}}
\def\arr{\begin{array}{rll}}
\def\ea{\end{array}}
\def\bea{\begin{eqnarray}}
\def\eea{\end{eqnarray}}

\def\N2{$N{=}2$}

\def\>{\rangle}
\def\<{\langle}
\def\+{\dagger}
\def\={\ =\ }

\begin{document}
\renewcommand{\thefootnote}{\fnsymbol{footnote}}
\begin{titlepage}
\setcounter{page}{0}
\vskip 1cm
\begin{center}
{\LARGE\bf  Eisenhart lift for higher derivative systems  }\\

\vskip 1.5cm

$
\textrm{\Large Anton Galajinsky and Ivan Masterov\ }
$
\vskip 0.7cm
{\it
Laboratory of Mathematical Physics, Tomsk Polytechnic University, \\
634050 Tomsk, Lenin Ave. 30, Russian Federation} \\
{Emails: galajin@tpu.ru, masterov@tpu.ru}

\end{center}
\vskip 1cm
\begin{abstract} \noindent
The Eisenhart lift provides an elegant geometric description of a dynamical system of second order in terms of
null geodesics of the Brinkmann--type metric. In this work, we attempt to generalize the Eisenhart method
so as to encompass higher derivative models. The analysis relies upon
Ostrogradsky's Hamiltonian. A consistent geometric description seems feasible
only for a particular class of potentials. The scheme is exemplified by the Pais--Uhlenbeck oscillator.
\end{abstract}

\vspace{0.5cm}

\end{titlepage}

\renewcommand{\thefootnote}{\arabic{footnote}}
\setcounter{footnote}0

\noindent
{\bf 1. Introduction}\\

\noindent
In classical mechanics several methods are known which provide a consistent geometric description of a second order dynamical system.
In general, the idea is to represent the equations of motion as geodesic equations in an appropriately chosen curved spacetime or to embed them into geodesics of a larger system in such a way that the dynamics of the extra degrees of freedom is fixed provided the evolution of the original model is known.  The Jacobi approach (see, e.g., \cite{Ong}) and
the Eisenhart lift \cite{Eis} \footnote{As originally formulated in \cite{Eis}, the Eisenhart lift had not received much attention by physicists and had soon fallen into oblivion.
The framework has been rediscovered in \cite{duv} in studying the geometry behind the Bargmann central extension of the Galilei group which paved the way for numerous physical applications.} seem to be the most popular methods of that kind. Given a dynamical system with $n$ degrees of freedom, the former operates with a Riemannian metric on an $n$--dimensional manifold, while the latter yields
a Brinkmann--type metric \cite{Br} of Lorentzian signature in an $(n+2)$--dimensional spacetime which is of interest in the general relativistic context.

In addition to the aesthetic appeal of the geometrization of dynamics, the Eisenhart lift provides an efficient means of studying hidden symmetries of spacetime. In general, such symmetries are associated with Killing tensors. In a series of recent works \cite{GHKW}--\cite{CGHHZ} various Lorentzian spacetimes admitting irreducible Killing tensors of rank greater than two have been constructed by applying the Eisenhart lift to specific integrable models. In \cite{cg,fg} Ricci--flat spacetimes of the ultrahyperbolic signature which support higher rank Killing tensors or possess maximally superintegrable geodesic flows have been built along similar lines.
Hidden symmetries of the Eisenhart lift metrics and the Dirac equation with flux have been studied in \cite{Marco}. Geometric uplifts of time-dependent systems were explored in \cite{cdgh}. An application to condensed matter physics was reported very recently in \cite{CGP}.

To the best of our knowledge, geometrizations of higher derivative systems of classical mechanics have not yet been studied in any detail. Although
higher derivative theories generically show up instability in classical dynamics and bring about violation of unitarity and/or trouble with ghosts in quantum theory, some of them, e.g. the Pais--Uhlenbeck oscillator \cite{PU}, are physically consistent and do deserve a geometric formulation. The goal of this work is to construct the Eisenhart lift
for a particular class of higher derivative models.

The paper is organized as follows. In Sect. 2 the original Eisenhart approach is reviewed with an emphasis on its Hamiltonian version. In Sect. 3 we analyze Ostrogradsky's Hamiltonian for the simplest class of
dynamical system of order $2n$, where $n$ is a natural number. It is demonstrated that, in view of the terms linear in momenta which are present in the Hamiltonian, the conventional Eisenhart procedure fails as it yields a degenerate metric. An alternative method, which operates with a larger set of extra degrees of freedom, is proposed and shown to yield a consistent geometric description for a particular class of potentials which
are the sum of homogeneous functions with arbitrary coefficients (coupling constants). Geometric properties of such metric are discussed in detail. The procedure is illustrated by the examples of the fourth order Pais--Uhlenbeck oscillator and its nonlinear generalization in \cite{Smilga}. In Sect. 4
an alternative possibility is considered which relies upon a simple canonical transformation applied to Ostrogradsky's Hamiltonian. It makes the conventional Eisenhart lift feasible, provided
the potential depends on the variable and its derivatives of even order only. The Pais--Uhlenbeck oscillator exemplifies the scheme. Sect. 5 contains the discussion and outlook.
\vspace{0.5cm}

\noindent
{\bf 2. Eisenhart lift for second order systems}\\

\noindent
The Eisenhart lift \cite{Eis} provides a geometric description of a dynamical system with $n$ degrees of freedom $x_1,\dots,x_n$ in terms of
null geodesics associated with the Brinkmann--type metric\footnote{For applications of the Brinkmann metric in other physical contexts see \cite{duv},\cite{dgh}--\cite{dhh}.} formulated in $(n+2)$--dimensional spacetime of Lorentzian signature
\be\label{metric}
d \tau^2=g_{AB}(z) d z^A d z^B=-2 U(x) d t^2+2 dt ds+\sum_{i=1}^n {(d x_i)}^2,
\ee
where $z^A=(t,s,x_1,\dots,x_n)$ and $U(x)$ is the potential which governs the dynamics of the original second order mechanics\footnote{For simplicity, we ignore possible interaction with external vector field potential $A_i(x)$ which would add the extra term $2 A_i(x) dt dx_i$ to the metric \cite{Br}.}.
Rewriting the null geodesic equations in components
\begin{align}\label{eqs1}
&
\frac{d^2 x_i}{d t^2}+\partial_i U(x)=0, &&  \frac 12 \sum_{i=1}^n {\left(\frac{d x_i}{d t}\right)}^2+U(x)=-c_2,
\nonumber\\[2pt]
&
\frac{d t}{d \tau}=c_1, &&  \frac{d s}{d t}-2 U(x)=c_2,
\end{align}
where $c_1$ and $c_2$ are constants of integration, one concludes that $t$ is affinely related to $\tau$, while $s$ decouples from the rest and its dynamics is unambiguously fixed provided the evolution of $x_i$ is known.  The original second order system is thus recovered by
implementing the null reduction along $s$ \cite{Eis}. A remarkable feature of the Eisenhart metric is that it admits the null and covariantly constant Killing vector field $\xi=\frac{\partial}{\partial s}$ which means that it belongs to the class of Kundt spacetimes.

An alternative possibility to construct the Eisenhart metric (\ref{metric}) is to start with the Hamiltonian corresponding to the original dynamical system
\be\label{HAM}
H=\frac 12 \sum_{i=1}^n p_i p_i+U(x),
\ee
where $(x_i,p_i)$ with $i=1,\dots,n$ form the canonical pairs, introduce two extra canonical pairs $(t,p^{(t)})$, $(s,p^{(s)})$ and promote (\ref{HAM}) to the specific function quadratic in momenta in the extended phase space \cite{GHKW}
\be\label{Hamm}
\tilde H=\frac 12 \sum_{i=1}^n p_i p_i+U(x) p^{(s)}p^{(s)}+p^{(s)}p^{(t)}.
\ee
Note that for mechanics interacting with external vector field potential terms linear in momenta are present in the original Hamiltonian. When constructing the extension, they should be multiplied by $p^{(s)}$  \cite{GHKW}.
It is easy to verify that the equations of motion following from (\ref{Hamm}) imply that $p^{(t)}$ and $p^{(s)}$ are constants of the motion while $t$ is affinely related to the evolution parameter $\tau$: $\frac{d t}{d \tau}=p^{(s)}$. Assuming $p^{(s)}\ne 0$ and switching from $\tau$ to $t$ in the remaining equations, one gets
\be
\frac{d^2 x_i}{d t^2}+\partial_i U(x)=0, \qquad \frac{d s}{dt}-2 U(x)=\frac{p^{(t)}}{p^{(s)}},
\ee
which reproduces the dynamical content of (\ref{eqs1}). Introducing the notation
\be
\tilde H=\frac 12 g^{AB}(z) P_A P_B,
\ee
where $P_A=(p^{(t)},p^{(s)},p_i)$, and considering $\tilde H$ as the geodesic Hamiltonian, one arrives at the Eisenhart metric (\ref{metric}). In this framework, the condition that the geodesic is null is usually interpreted as the fact that the time translation generator $\partial_t$ in the spacetime is linked to the Hamiltonian governing the dynamics of the original system (\ref{HAM}).

\vspace{0.5cm}

\noindent
{\bf 3. Eisenhart lift for higher derivative models via Ostrogradsky's Hamiltonian}\\

\noindent
Consider a particular class of dynamical systems of order $2n$ for which the highest derivative is separated from the rest in the Lagrangian
\be\label{Lag}
L=\frac 12 x^{(n)} x^{(n)}-U\left(x,\dot x, \dots, x^{(n-1)}\right),
\ee
where $\dot x=\frac{d}{d t} x(t)$, $x^{(k)}=\frac{d^k}{d t^k} x(t)$ and $t$ is the evolution parameter.
The corresponding equation of motion reads
\be\label{EM}
x^{(2n)}+\sum_{k=0}^{n-1} {(-1)}^{k+n+1} \frac{d^k}{d t^k} \left( \frac{\partial }{\partial x^{(k)}} U\left(x,\dot x, \dots, x^{(n-1)}\right) \right)=0.
\ee
In the next section we shall consider a more general Lagrangian involving also the linear contribution $-x^{(n)}V\left(x,\dot x, \dots, x^{(n-1)}\right)$.

As was mentioned in the preceding section, a conventional means of constructing the Eisenhart metric associated with a second order dynamical system is to extend its phase space by the extra canonical pairs $(t,p^{(t)})$, $(s,p^{(s)})$ and promote the Hamiltonian to a specific function quadratic in momenta which determines the inverse Eisenhart metric.

The standard Hamiltonian formulation for the higher derivative system (\ref{Lag}) is built with the use of Ostrogradsky's method
\be\label{Ham}
H=\frac 12 p_n^2+\sum_{\alpha=1}^{n-1} p_\alpha x_{\alpha+1}+U(x_1,\dots,x_n),
\ee
where the variables $(x_n,p_n)$ and $(x_\alpha,p_\alpha)$ with $\alpha=1,\dots,n-1$ form the canonical pairs and $x_1$ is identified with the original dynamical variable $x$ in (\ref{EM}).
In particular, the equations of motion following from (\ref{Ham}) include the chain of relations
\be\label{xrel}
{\dot x}_\alpha=x_{\alpha+1}.
\ee
As far as a putative geometric formulation of the Hamiltonian system (\ref{Ham}) is concerned, the first order relations (\ref{xrel}) reveal a subtlety. Because geodesic equations are of the second order,
(\ref{xrel}) should arise as first integrals. However, a generic first integral involves a constant of integration. It is thus likely that within the Eisenhart--like approach to the geometrization of higher derivative systems Eq. (\ref{xrel}) should be modified so as to include arbitrary constants. The resulting geometric formulation will encompass a larger class of models only a particular member of which will reproduce the dynamical system (\ref{Ham}). Below we discuss a variant of the Eisenhart lift for which (\ref{xrel}) is promoted to the first integrals of the form
\be
\frac{{\dot x}_\alpha}{x_{\alpha+1}}=C_\alpha,
\ee
where $C_\alpha$ are arbitrary constants. For the extended dynamical system these are interpreted as coupling constants.

An attempt to construct the conventional Eisenhart metric associated with the Hamiltonian (\ref{Ham}) reveals a problem. The metric turns out to be degenerate. In order to circumvent the difficulty, let us extend Ostrogradsky's phase space by a set of extra variables $(t,p^{(t)})$, $(s_\alpha,p^{(s)}_{\alpha})$ with $\alpha=1,\dots,n-1$ and introduce the Hamiltonian which governs the dynamics in the extended phase space
\be\label{tH}
\tilde H=\frac 12 p_n^2+\sum_{\alpha=1}^{n-1} p^{(s)}_{\alpha} p_\alpha x_{\alpha+1}+U(x_1,\dots,x_n)\sum_{\alpha=1}^{n-1} p^{(s)}_{\alpha} p^{(s)}_{\alpha}+\frac 12 p^{(t)} p^{(t)}.
\ee
As follows from (\ref{tH}), $p^{(t)}$ and $p^{(s)}_{\alpha}$ are constants of the motion while the evolution of $s_\alpha$ is fixed provided the general solution to the equations of motion for the original phase space variables is known.
The dynamics of the sector $(x_n,p_n)$, $(x_\alpha,p_\alpha)$ is thus split from the evolution of the extra variables $(t,p^{(t)})$, $(s_\alpha,p^{(s)}_{\alpha})$
which is one of the key features of the Eisenhart lift.

The Eisenhart--like metric associated with the Hamiltonian (\ref{tH}) reads
\be\label{EMetr}
d\tau^2=g_{AB}(z) dz^A dz^B=dt^2+dx_n^2+2\sum_{\alpha=1}^{n-1} \frac{d x_\alpha d s_\alpha}{x_{\alpha+1}}-2 U(x_1,\dots,x_n) \sum_{\alpha=1}^{n-1} \frac{d x_\alpha^2}{x_{\alpha+1}^2},
\ee
where $z^A=(t,s_\alpha,x_\alpha,x_n)$, $\alpha=1,\dots,n-1$, $A=1,\dots,2n$. Introducing the geodesic Lagrangian $L=\frac 12 g_{AB} (z) \dot z^A \dot z^B$, where $\dot z^A=\frac{d z^A}{d \tau}$ and $\tau$ is the proper time, and adopting the notation
\be\label{pi}
\pi_\alpha=\frac{1}{x_{\alpha+1}}\left({\dot s}_\alpha  -2 U \frac{\dot x_\alpha}{x_{\alpha+1}} \right),
\ee
one obtains the geodesic equations
\bea\label{eqs}
&&
\ddot t=0, \qquad \qquad \qquad \qquad \qquad \qquad \qquad \qquad
{\left( \frac{{\dot x}_\alpha}{x_{\alpha+1}}\right)}^{\cdot}=0,
\\[2pt]
&&
{\ddot x}_n+\pi_{n-1} \frac{{\dot x}_{n-1}}{x_n}+\partial_n U \sum_{\beta=1}^{n-1} {\left(\frac{{\dot x}_\beta}{x_{\beta+1}} \right)}^2=0, \quad
{\dot\pi}_\alpha+\pi_{\alpha-1} \frac{{\dot x}_{\alpha-1}}{x_\alpha}+\partial_\alpha U \sum_{\beta=1}^{n-1} {\left(\frac{{\dot x}_\beta}{x_{\beta+1}} \right)}^2=0,
\nonumber
\eea
where it is assumed that $x_0=\pi_0=0$ and $\partial_\alpha U=\frac{\partial U}{\partial x_\alpha}$, $\partial_n U=\frac{\partial U}{\partial x_n}$.

The first line in Eq. (\ref{eqs}) implies that $t$ is affinely related to the proper time $\tau$ while $\frac{{\dot x}_\alpha}{x_{\alpha+1}}$ are constants of the motion
\be
\frac{{\dot x}_\alpha}{x_{\alpha+1}}={C}_\alpha.
\ee
These relations generalize (\ref{xrel}). In what follows we consider all ${C}_\alpha$ to be nonzero and abbreviate
\be
\Omega=
\sum_{\beta=1}^{n-1} C_\beta^2.
\ee
For $\alpha=2,\dots,n-1$ the rightmost equation entering the second line in (\ref{eqs}) yields the recurrence relation which links $\pi_{\alpha-1}$ to ${\dot\pi}_{\alpha}$ and $\partial_{\alpha} U$, while the leftmost equation in the second line of (\ref{eqs}) fixes $\pi_{n-1}$ in terms of ${\ddot x}_n$ and $\partial_n U$. Given the definition of $\pi_\alpha$ in (\ref{pi}), one concludes that all together these equations provide a set of the first order differential equations which unambiguously fix $s_\alpha$, provided the dynamics of $x_\alpha$ and $x_n$ is known. The remaining equation $\dot\pi_1+\partial_1 U \Omega=0$ yields
\be\label{final}
x^{(2n)}+[{(C_1\dots C_{n-1})}^2 \Omega] \sum_{k=0}^{n-1} {(-1)}^{k+n+1} \frac{d^k}{d \tau^k} \left( \frac{\partial }{\partial x^{(k)}} U\left(x,\frac{\dot x}{C_1} , \dots, \frac{ x^{(n-1)}}{C_1 \dots C_{n-1}}\right) \right)=0,
\ee
where we denoted $x_1=x$. Comparing Eqs. (\ref{final}) and (\ref{EM}), one concludes that the geodesics of the Eisenhart--like metric (\ref{EMetr}) describe an $(n-1)$--parametric deformation of
the original dynamical system (\ref{EM}), $C_\alpha$ being the deformation parameters. Note that for systems of the fourth order all factors including $C_1$ can be removed by redefining the proper time $C_1 \tau \to \tilde\tau$ such that (\ref{final}) reduces exactly to (\ref{EM}), while for generic potentials such rescaling gives $U\left(x,\dot x,\frac{C_1}{C_2} \ddot x, \dots, \left[\frac{ C_1}{C_2} \dots \frac{ C_1}{C_{n-1}}\right]  x^{(n-1)}\right)$.

If the factors
${(C_1\dots C_{n-1})}^2 \Omega$ and $\frac{1}{C_1}, \dots, \frac{1}{C_1 \dots C_{n-1}}$ in (\ref{final}) can be removed by redefining coupling constants entering the original potential,
the deformation is fictitious and the metric (\ref{EMetr}) provides a valid geometric description of (\ref{EM}).
In particular, this occurs for potentials of the form
\bea
&&
U(x_1,x_2,\dots,x_n)=\sum_{i=1}^N g_i W_i (x_1,x_2,\dots,x_n),
\eea
where $W_i(x_1,x_2,\dots,x_n)$ are homogeneous functions of the arguments $x_3,\dots,x_n$ of possibly different degrees $k_i$
\be
W_i(x_1,x_2,\lambda x_3,\dots,\lambda x_n)=\lambda^{k_i} W_i(x_1,x_2,x_3,\dots,x_n)
\ee
and $g_i$ are coupling constants.

As an example, let us consider the fourth order Pais--Uhlenbeck oscillator which is described by the Lagrangian
\be\label{LPU}
L=\frac 12 {\ddot x}^2-\frac 12 (\omega_1^2+\omega_2^2) {\dot x}^2+\frac 12 \omega_1^2 \omega_2^2 x^2,
\ee
where $\omega_1$ and $\omega_2$ are two distinct frequencies of oscillation, and the equation of motion
\be\label{PUem}
\left(\frac{d^2}{dt^2}+\omega_1^2 \right) \left(\frac{d^2}{dt^2}+\omega_2^2 \right)x=0.
\ee
In this case the metric (\ref{EMetr}) takes the form
\bea
&&
d \tau^2=dt^2+d x_2^2+\frac{2}{x_2} d x_1 d s_1-\left(\omega_1^2+\omega_2^2-  \omega_1^2 \omega_2^2 {\left(\frac{x_1}{x_2}\right)}^2\right) d x_1^2,
\eea
while the geodesic equations include
\be
x_1^{(4)}+C_1^2 (\omega_1^2+\omega_2^2) \ddot x_1+C_1^4 \omega_1^2 \omega_2^2 x_1=0,
\ee
where $\dot x_1=\frac{d x_1}{d \tau}$ and $\tau$ is the proper time. Redefining the evolution parameter $C_1 \tau \to \tilde\tau$ one reproduces (\ref{PUem}). Alternatively one can rescale the frequencies $C_1 \omega_{1,2} \to \tilde\omega_{1,2}$.

One more example is given by a nonlinear system introduced by Smilga in studying the stability of higher derivative mechanics \cite{Smilga}
\be
L=\frac 12 {\left(\ddot x+\omega^2 x \right)}^2-\frac{\alpha}{4} x^4-\frac{\beta}{2} x^2 {\dot x}^2,
\ee
where $\omega$, $\alpha$ and $\beta$ are arbitrary constants. Its geometrization is given by (\ref{EMetr}) which involves
\be
U(x_1,x_2)=-\frac{\omega^4}{2} x_1^2+\omega^2 x_2^2+\frac{\alpha}{4} x_1^4+\frac{\beta}{2} x_1^2 x_2^2.
\ee

Geometric description of higher derivative systems reveals properties which are strikingly different from those characterizing second order models.  The metric (\ref{EMetr}) is of the ultrahyperbolic signature\footnote{For $n=2$ the signature is Lorentzian. Yet, the spacetime is parametrized by three temporal and one spatial coordinates.}. The null Killing vector fields $\xi^{(\alpha)}=\frac{\partial}{\partial s_\alpha}$ fail to be covariantly constant. The spacetime has curvature singularities along the hyperplanes $x_\alpha=0$ with $\alpha=2,\dots,n$. Irrespective of the explicit form of the potential $U(x_1,\dots,x_n)$, (\ref{EMetr}) does not solve the vacuum Einstein equations. This is to be contrasted with (\ref{metric}) which is Ricci--flat provided the potential is a harmonic function \cite{dgh}, while for spacetimes of the ultrahyperbolic signature the potential should be an additive function \cite{cg}. In view of the ultrahyperbolic signature, it proves problematic to unambiguously link the Hamiltonian of the original mechanics to the time translation generator in spacetime. Thus, the specification to null geodesics conventionally adopted for second order systems seems to be superfluous.

\vspace{0.5cm}

\noindent
{\bf 4. Canonical transformation of Ostrogradsky's Hamiltonian and Eisenhart lift}\\

\noindent
As was mentioned in the preceding section, a naive treatment of Ostrogradsky's Hamiltonian within the Eisenhart framework yields a degenerate metric. The problem is rooted in terms linear in momenta which are
present in the Hamiltonian. As is well known, higher derivative dynamics may admit more than one Hamiltonian description (see, e.g., Refs. \cite{AGMM,EG} and references therein). In this section we consider an alternative possibility which consists in applying the simple canonical transformation
\bea\label{canon}
&&
x_{2k}\rightarrow p_{2k},\quad p_{2k}\rightarrow -x_{2k},
\eea
with $k=1,\dots,[\frac{n}{2}]$, which removes the unwanted linear terms entering the kinetic part provided the original potential depends on $x$ and its derivatives of even order only.
The transformed system turns out to be the conventional mechanics in pseudo--Euclidean space to which the original Eisenhart lift can be straightforwardly applied.
For what follows it proves convenient to treat the cases of even and odd values of $n$ separately and consider a more general Lagrangian which also involves the contribution linear in the highest derivative
\bea\label{HDI}
L=\frac{1}{2}x^{(n)}x^{(n)}-x^{(n)}V\left(x,\ddot{x},x^{(4)},\dots,\epsilon x^{(n-1)}\right)-U\left(x,\ddot{x},x^{(4)},\dots,\epsilon x^{(n-1)}\right),
\eea
where $\epsilon=1$ for even $(n-1)$ and $\epsilon=0$ for odd $(n-1)$.

For $n=2m$ the equation of motion reads
\bea\label{equ}
x^{(4m)}-\frac{d^{2m}V}{dt^{2m}}-\sum_{k=0}^{m-1}\frac{d^{2k}}{dt^{2k}}\left(x^{(2m)}\frac{\partial V}{\partial x^{(2k)}}+\frac{\partial U}{\partial x^{(2k)}}\right)=0.
\eea
Constructing Ostrogradsky's Hamiltonian and implementing the canonical transformation (\ref{canon}), one gets
\bea\label{Halt}
H'=\sum_{k=1}^{m}p_{2k-1}p_{2k}+W(x),\qquad W(x)=\frac 12 {\left(V-x_{2m} \right)}^2+U-\sum_{k=1}^{m-1}x_{2k}x_{2k+1},
\eea
where $U=U(x_1,x_3,\dots,x_{2m-1})$ and $V=V(x_1,x_3,\dots,x_{2m-1})$. It is straightforward to verify that the canonical equations of motion resulting from (\ref{Halt}) do reproduce (\ref{equ}).
Because (\ref{Halt}) is formulated as the conventional mechanics in pseudo--Euclidean space, the standard Eisenhart extension is feasible
\bea\label{H1}
\tilde H=\sum_{k=1}^{m}p_{2k-1}p_{2k}+W(x) p^{(s)} p^{(s)}+p^{(s)} p^{(t)},
\eea
which yields the metric
\bea\label{MM}
d\tau^2=-W(x)dt^2+dt ds+\sum_{k=1}^{m}dx_{2k-1}dx_{2k},
\eea
where $W(x)$ is given in (\ref{Halt}).
The geodesic equations associated with (\ref{MM}) do reproduce (\ref{equ}), while the evolution of $s$ is fixed provided the general solution of (\ref{equ}) is known. For earlier application of the Eisenhart lift to mechanics in pseudo--Euclidean space see \cite{cg,fg}.

Turning to the odd values of $n=2m+1$, the condition that the function $x^{(n)}V+U$ depends on $x$ and its derivatives of even order only implies
\bea
V=0,\qquad U=U\left(x,\ddot{x},..,x^{(2m)}\right),
\eea
while the equation of motion reads
\bea\label{equ2}
x^{(4m+2)}+\sum_{k=0}^{m}\frac{d^{2k}}{dt^{2k}}\frac{\partial U}{\partial x^{(2k)}}=0.
\eea
After performing the canonical transformation (\ref{canon}), Ostrogradsky's Hamiltonian associated with Eq. (\ref{equ2}) takes the
form\footnote{The Hamiltonians (\ref{Halt}) and (\ref{Halt1}) can be also derived from the Lagrangian (\ref{HDI}) by the method in \cite{AGMM}.}
\bea\label{Halt1}
\begin{aligned}
H'=\frac{1}{2}p_{2m+1}^2+\sum_{k=1}^{m}p_{2k-1}p_{2k}+W(x),\qquad W(x)=U(x_1,x_3,..,x_{2m+1})-\sum_{k=1}^{m}x_{2k}x_{2k+1},
\end{aligned}
\eea
which gives rise to the extended Hamiltonian and the Eisenhart metric
\bea
&&
\tilde H=\frac{1}{2}p_{2m+1}^2+\sum_{k=1}^{m}p_{2k-1}p_{2k}+W(x) p^{(s)} p^{(s)}+p^{(s)} p^{(t)}
\nonumber\\[2pt]
&&
d\tau^2=-W(x) dt^2+ds dt+\sum_{k=1}^{m}dx_{2k-1}dx_{2k}+\frac 12 dx_{2m+1}^2.
\eea

As far as applications are concerned, the method above fits perfectly to geometrize the celebrated Pais--Uhlenbeck oscillator \cite{PU} which is characterized by the potentials\footnote{We use the notation in \cite{AGGM}.}
\bea
V=-\frac{1}{2}\sum_{k=m}^{2m-1}\sigma^{n}_{k} x^{(2k-2m)}, \quad U=-\frac{1}{2} x \sum_{k=0}^{m-1}\sigma^{n}_{k}x^{(2k)}, \quad
\sigma_k^n=\sum_{i_1<i_2<..<i_{n-k}}^{n}\omega_{i_1}^2 \dots \omega_{i_{n-k}}^2,
\eea
for even $n=2m$ or
\bea
U=\frac{1}{2} x^{(2m)} \sum_{k=0}^{m}\sigma^{n}_{m+k}x^{(2k)}+\frac{1}{2} x \sum_{k=0}^{m-1}\sigma_k^{n}x^{(2k)}
\eea
for odd $n=2m+1$.

Like in the preceding section, the Eisenhart metrics constructed above are of the ultrahyperbolic signature. This is because Eqs. (\ref{Halt}) and (\ref{Halt1}) appeal to mechanics in pseudo--Euclidean space.
Worth mentioning also is that, within the alternative Hamiltonian formulation adopted in this section, the analogues of the first order relations (\ref{xrel}) read
\be
\frac{d^2 x_{2k-1}}{dt^2}-x_{2k+1}=0.
\ee
Being the second order equations, these fit perfectly to be embedded into the geodesic equations associated with the conventional Eisenhart metric.

\vspace{0.5cm}

\noindent
{\bf 5. Conclusion}\\

\noindent
To summarize, in this work a possibility to generalize the Eisenhart lift so as to encompass higher derivative systems was examined. The analysis relied upon a proper extension of
Ostrogradsky's Hamiltonian formulation. A consistent geometric description proved feasible only for a particular class of potentials. It includes potentials which are the
sum of homogeneous functions with arbitrary coefficients (coupling constants) or depend on the variable and its derivatives of even order only. The consideration was exemplified by
the Pais--Uhlenbeck oscillator.

A number of interesting issues deserve a further consideration. The metrics constructed in this work are of the ultrahyperbolic signature. Although this seems to be an indispensable feature, it is interesting to understand whether Lorentzian spacetimes may be associated with higher derivative systems by developing alternative approaches. An important issue is to study how global symmetries of the original higher derivative mechanics are transmitted into those of the Eisenhart metric. As was mentioned in Sect. 3, a straightforward attempt to construct the Eisenhart metric associated with the conventional Ostrogradsky' Hamiltonian yields a degenerate metric tensor. It would be interesting to investigate whether an analogue of the Newton--Cartan geometry, which operates with a divergent metric, can be developed in this case.
A consistent geometrization of generic unconstrained potentials remains a challenge.

\vspace{0.5cm}

\noindent{\bf Acknowledgements}\\

\noindent
We thank Peter Horv\'athy for useful comments.
This work was supported by the MSE program Nauka under the project 3.1113.2017/Pp, 
the RFBR grant 17-02-00047, and the RF Presidential grant MK-2101.2017.2.

\end{document}